\documentclass[english]{article}
\usepackage[T1]{fontenc}
\usepackage[latin9]{inputenc}  
\usepackage{textcomp}
\usepackage{amsmath}
\usepackage{amssymb}
\usepackage{graphicx}
\usepackage{esint}

\makeatother
\usepackage{array}
\usepackage{textcomp}
\usepackage{multirow}
\makeatletter
\usepackage{babel}

\author{
\scshape P. Lautridou\footnote{Proceeding of the ARENA 2012 conference}~~and the CODALEMA collaboration\\
\small \textit{SUBATECH, Univ. de Nantes-Ecole des Mines de Nantes-CNRS,}\\ 
\small \textit{4 rue A. Kastler, 44307 Nantes, France}\\
\small \textit{LESIA, USN de  Nançay, Obs. de Paris-Meudon-CNRS,} \\
\small \textit{5 place J. Janssen, 92195 Meudon, France}
}

\title{\textbf{Some possible interpretations from data of the CODALEMA experiment}} 

\begin{document}
\maketitle

\begin{abstract}
The purpose of the CODALEMA experiment, installed at the Nançay Radio Observatory (France), is to study the radio-detection of ultra-high energy cosmic rays in the energy range of $10^{16}-10^{18}~eV$. Distributed over an area of 0.25 km$^2$, the original device uses in coincidence an array of particle detectors and an array of short antennas, with a centralized acquisition. A new analysis of the observable in energy for radio is presented from this system, taking into account the geomagnetic effect. Since 2011, a new array of radio-detectors, consisting of 60 stand-alone and self-triggered stations, is being deployed over an area of 1.5 km$^2$ around the initial configuration. This new development leads to specific constraints to be discussed in term of recognition of cosmic rays and in term of analysis of wave-front.
\end{abstract}

\begin{center}
\textbf{keywords:} Ultra-High Energy Cosmic Rays; Radio-detection; Energy; Wavefront
\end{center}

\section{Introduction}

The radio-detection of ultra-high energy cosmic rays (UHECR) may be considered as one of the future tools, likely to supplement the traditional measures of extensive air showers (EAS) \cite{UHECR12}. This method exploits, through antennas, the radio signal that accompanies the development of the EAS. Many results have highlighted its feasibility and its potential in reconstructing several parameters, as the arrival direction, the impact location, or the radio lateral distribution function of the electric field (RLDF) \cite{ard06, apel10, mar11}. However, except a recent study \cite{gla12}, the estimation of the energy through radio has been only briefly discussed \cite{horn07, rav12}, although this step is unavoidable if one really plans to use this technique. Similarly, the radio wavefront characteristic is remained poorly determined \cite{laf10, trend11, weid12}, although its knowledge could be consider as one of the first steps in retrieving information about the EAS itself. These issues motivate the development and the continuation of demonstrative  experiences, such as CODALEMA \cite{nim05}.\\
This paper summarizes some new aspects, especially regarding the energy estimator \cite{Arxivcor} and the point source of the radio emission \cite{Arxivill}. 

\section{Experimental setups}

\begin{figure}[htp]
\centering
\includegraphics[width=8cm,height=6cm]{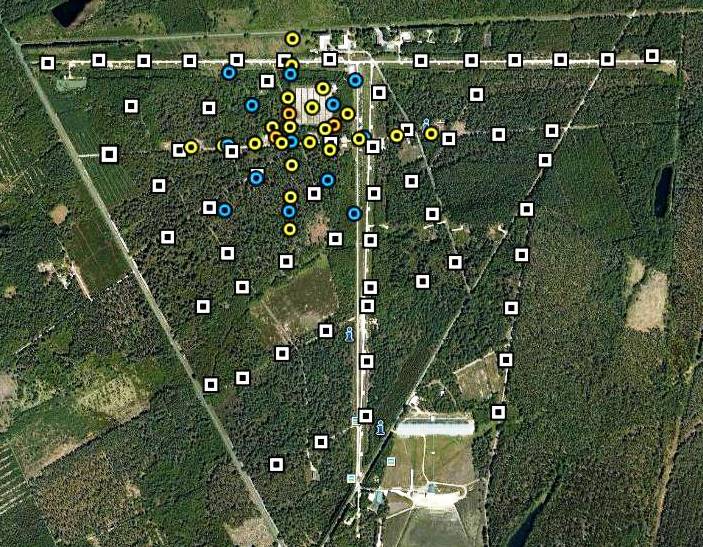}
\caption{Aerial view of the CODALEMA setup installed at Nan\c cay. On the original system, the plastic scintillators are represented by blue circles while yellow and orange circles represent the dipole antennas, placed respectively in the $EW$ and $NS$ directions. The white squares show the locations of the autonomous stations of the new array in deployment (33 are currently in operation).}
\label{Setup}
\end{figure}

The study of the energy response in radio took advantage of the original setup which uses a centralized trigger (see \cite{ARENA08}). It consists of two arrays of detectors (fig. \ref{Setup}): an antenna array of $24$ short active dipoles distributed on cross of $600\times 500~m$; a particle detector array of $17$ plastic scintillator plates spread over a square grid of $340\times 340~m$. The antennas operate in the frequency bandwidth $1-200~MHz$. The scintillating detectors are read by 2 photomultipliers (one at high gain, the other at low gain) so as to cover an overall dynamics of 0.3 to 3000 vertical equivalent muons (VEM).
 
The trigger system is provided by the coincident detection of the five central particle detectors, leading to a detection threshold around $10^{15.7}~eV$. For each trigger, all the sensors signals are digitized at a sampling rate of $1~GS/s$ (snapshot of $2.56~\mu s$).\\
During the off-line analysis, the radio transient is searched in the radio waveform through the $24-82~MHz$ bandwidth, from which the electric field peak value and the arrival time are extracted. If more than 3 radio transients are detected in coincidence, the arrival direction of the electric field (zenith angle $\theta $, azimuth angle $\phi $) is reconstructed by considering a planar wavefront. Assuming a RLDF for the electric field of the form $\varepsilon_{0}~\exp(-d/d_{0})$ \cite{all71}, four additional observables are derived: the electric field magnitude at the radio shower center $\varepsilon_{0}$, the slope parameter of the electric field $d_{0}$, and the impact location of the radio shower center on the ground ($x_{0},y_{0}$). The standard deviation of the noise on each antenna is used as the error on the electric field measured. From Monte-Carlo simulations, an mean error of $22~\%$ is estimated for $\varepsilon_{0}$. The timing error is evaluated to $10~ns$, leading to an angular accuracy of reconstructed arrival directions in the range of the degree.\\
Meanwhile, the arrival direction of the particle front (assumed planar) is computed from the relative arrival time of each particle detector and an analytical NKG lateral distribution \cite{key-28} is adjusted on the measured particle densities in the shower frame. If the shower core position falls inside the particle detector array, the event is referred to as ``internal'' and a adjusted NKG distribution is used to deduce the primary energy using the constant intensity cut (CIC) method \cite{key-29}. This processing leads to a ``particle'' energy resolution of the order of $30~\%$ at $10^{17}~eV$.\\  
A coincidence below $ 100~ns$ in time and $20^{\circ }$ in arrival direction with the event ``particles'' labels the event ``radio'' as an EAS. Imposing zenith angles less than $60^{\circ }$, and energies ``particle'' above $10^{16.7}~eV$, $376$ coincidences were available for studies of the energy ``radio'', discussed in section 3. \\

The study of the wavefront features has been based on the new autonomous setup made of self-triggered antenna stations (see \cite{RICAP11}). The later are based both on the butterfy antenna and on a embedded electronic. Its aim is to complete the initial configuration, providing finer measurements of the electric field (with two polarization states per station: $EW$ and $NS$), better SNR, and an increased range in energy, using a larger area. 33 stations are currently in operation around the original setup. Ultimately, this apparatus will cover an area of $1.5~km^2$, with the additional aim to bring this technique reliable and mature for large infrastructure. Up to now, efforts are  mostly focused on the radio-sources, observed using the self-triggering technique. The interpretations of these observations are central to the fourth  section. The event are built from the time coincidences between the stations using the GPS time.

\section{Energy estimation}

\begin{figure}
\centering
\includegraphics[width=8cm,height=6cm]{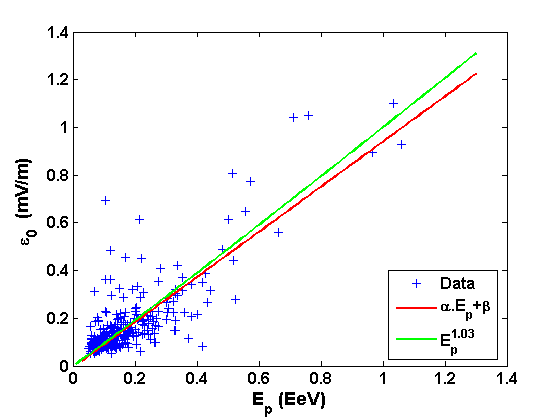}
\caption{Electric field $\varepsilon_{0}$ versus the primary energy $E_{P}$ (for readability, the error bars of $30$ \% for $E_{P}$ and $22$ \% for $\varepsilon_{0}$ are not represented). The result of the adjustment of the data with the power law $E_{P}^{1.03}$ has motivated the use of a linear adjustment depicted in red.}
\label{Raw}
\end{figure}

This study has focused on the possibility of deriving from radio, a simple observable which could provide a preliminary satisfactory estimate of the energy of the shower, by going further in some previous works \cite{rav12, ARENA08}. Because the electric field radiated should be linked to the energy of the primary particle through the number of secondary charges produced within the shower, we chose to exploit the maximum electric field $\varepsilon_{0}$ (derived from the RLDF) as radio-observable of the energy. For this, the correlation with the shower energy estimator $E_{P}$ (deduced from the CIC analysis) is exploited (fig. \ref{Raw}) by fitting the scatter plot distribution $(E_{P},~\varepsilon_{0})$ with the linear relation  $\varepsilon_{0}=\alpha\, E_{P}+\beta $.   

\begin{figure}
\centering
\includegraphics[width=8cm,height=6cm]{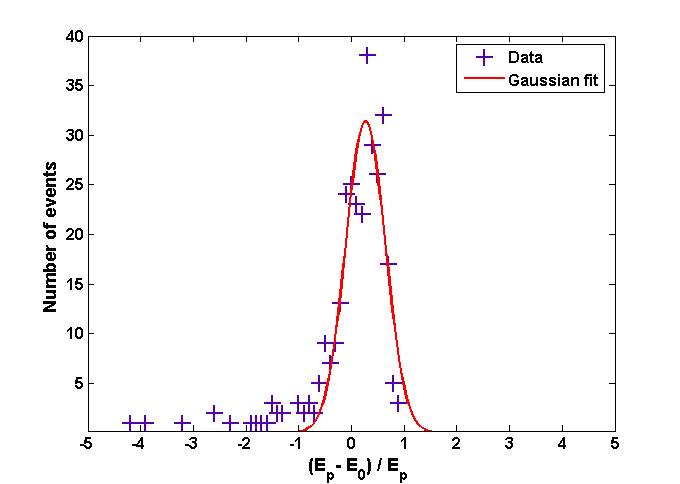}
\caption{Distribution $(E_{P}-E_{0})/E_{P}$ obtained after calibration and for the full sample of events
(blue crosses). The standard deviation is $31$ \% with $c=0.95$ (see below). The red line shows a Gaussian fit made in the interval $[-1,1]$.}
\label{Diff}
\end{figure}

Inversion of the latter gives the energy estimator of the radio $E_{0}=\varepsilon_{0}/\alpha-\beta/\alpha$. 
Discrepancies between both observables $E_{0}$ and $E_{P}$ is quantified through the analysis of the distribution of their relative differences $ (E_{P}-E_{0})/E_{P}$ characterized by its standard deviation (SD). 
A typical forms of the distribution $(E_{P}-E_{0})/E_{P}$ is presented in fig. \ref{Diff}.   

\begin{figure}
\centering
\includegraphics[width=8cm,height=6cm]{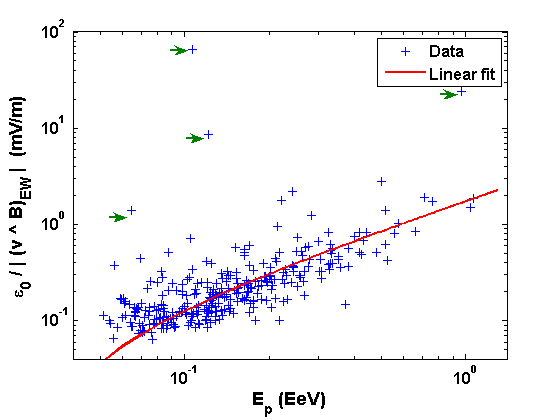}
\caption{Scatter plot of the corrected electric field $\varepsilon_{0}/|(\mathbf{v}\wedge\mathbf{B})_{EW}|$ versus
the primary energy $E_{P}$ . The linear adjustment is shown in red.
The green arrows show some of the typical over-corrected events. The
systematic analysis of their features indicates that all these events
have arrival directions near the earth's magnetic field direction
($|(\mathbf{v}\wedge\mathbf{B})_{EW}|<0.1$). (These points
are removed for the linear fit of the scatter plot).}
\label{VXB}
\end{figure}

\begin{figure}
\centering
\includegraphics[width=8cm,height=6cm]{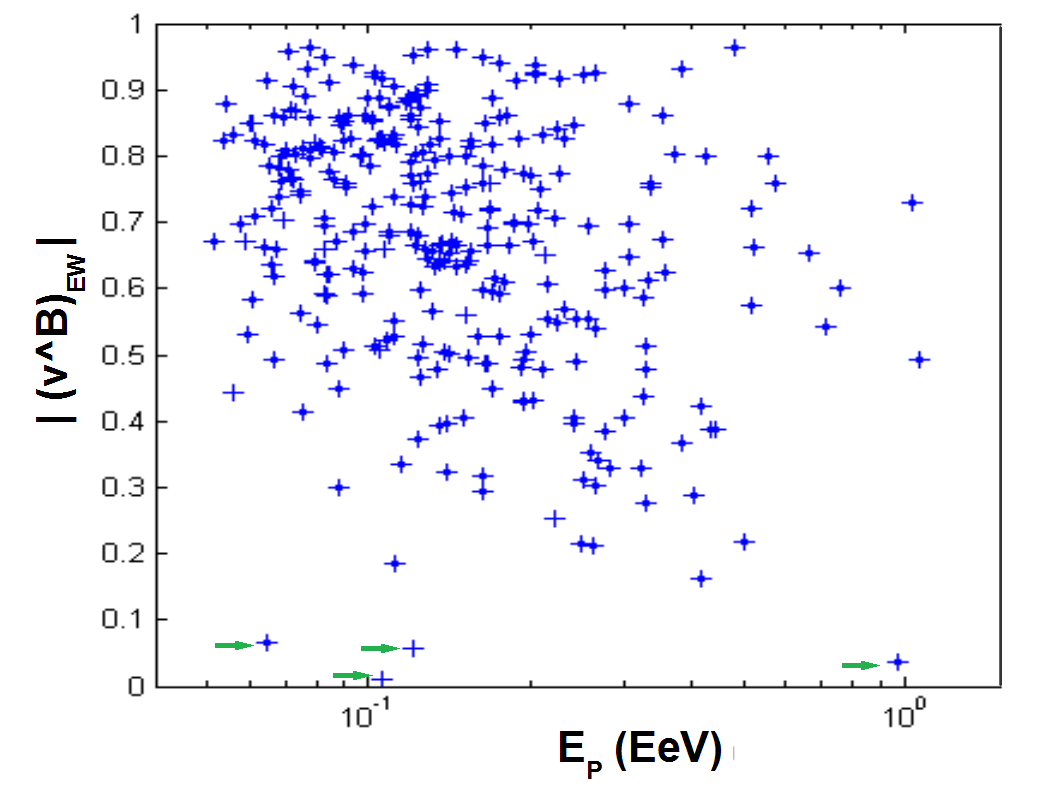}
\caption{Distribution of $|(\mathbf{v}\wedge\mathbf{B})_{EW}|$ versus
the primary energy $E_{P}$ for the raw sample of events.}
\label{EpVXB}
\end{figure}

To accord with our earlier works \cite{ard09}, which showed that the electric field magnitude depends on the Lorenz force via the vector cross product $\mathbf{v}\wedge\mathbf{B}$, we have corrected the
estimator $\varepsilon_{0}$ by the projection of the Lorentz force along the $EW$ polarization. So, two detected showers of the same actual primary energy $E$ (and of the same nature) will produce two different radio signals depending on
their angle with the geomagnetic axis, by following the relation:   $\varepsilon_{0}=E/|(\mathbf{v}\wedge\mathbf{B})_{EW}|$. 
In doing so, the scatter plot $(E_{P},~\varepsilon_{0}/|(\mathbf{v}\wedge\mathbf{B})_{EW}|)$ (fig. \ref{VXB}) shows that events located close to the earth's magnetic field direction ($|(\mathbf{v}\wedge\mathbf{B})_{EW}|<0.1$) gain unrealistic energies, while their CIC energies show nothing unusual. The original distribution of $(|(\mathbf{v}\wedge\mathbf{B})_{EW}|$ versus $E_P$ is presented fig \ref{EpVXB}. 

\begin{figure}
\centering
\includegraphics[width=8cm,height=6cm]{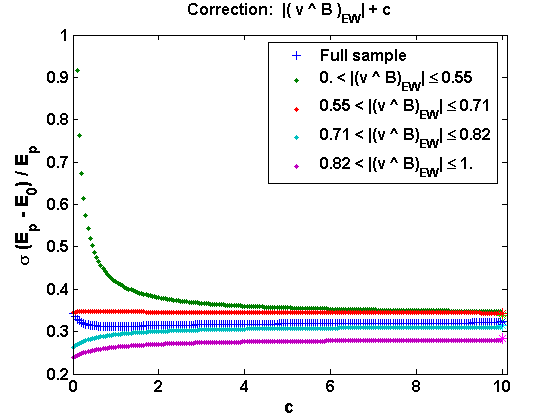}
\caption{Evolution of the Standard Deviation of the distribution $(E_{P}-E_{0})/E_{P}$ with the correction factor $1/(|(\mathbf{v}\wedge\mathbf{B})_{EW}|+c)$ as a function of $c$ and for several windows of $|(\mathbf{v}\wedge\mathbf{B})_{EW}|$.
The blue line represents the whole sample of events. The stars located at 
$c=10$ indicate the values of the SD without correction. The SD errors are given by $\sqrt{2}\:\sigma^{2}/\sqrt{N-1}$ ($N\approx70$ for the subsets).}
\label{VXBSD}
\end{figure}

To account for this, we hypothesized that the observed
electric field $\varepsilon_{0}$ can also be proportional to the actual shower energy $E$ (to produce observable signals, especially when the Lorentz contribution becomes weak). This leads to modify $\varepsilon_{0}$ toward $\varepsilon_{0}\rightarrow\varepsilon_{0}/(|(\mathbf{v}\wedge\mathbf{B})_{EW}|+c)$ where $c$ is a scalar quantity. 
With the quantity $|(\mathbf{v}\wedge\mathbf{B})_{EW}|$ ranging from 0 to 1, $c=0$ corresponds to a purely geomagnetic contribution, while a high $c$ value (e.g. $c>6$) should represent the case for which the geomagnetic emission becomes negligible compared to the second effect. A high $c$ value must induce a behavior similar to that observed without correction. The figure \ref{VXBSD} presents the standard deviation of $(E_{P}-E_{0})/E_{P}$ as a function of $c$ and for several bins in $|(\mathbf{v}\wedge\mathbf{B})_{EW}|$.

\begin{figure}
\centering
\includegraphics[width=8cm,height=6cm]{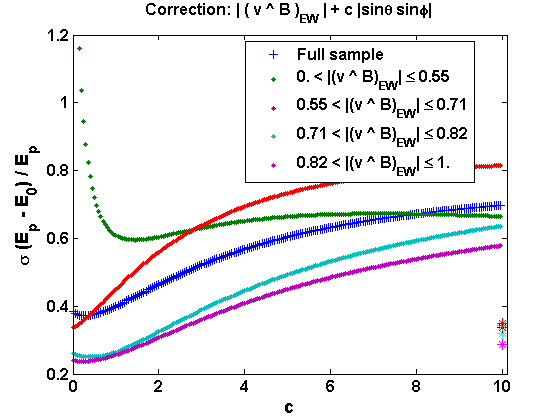}
\caption{Evolution of the Standard Deviation of the distribution $(E_{0}-E_{P})/E_{P}$
with the correction factor $1./(|(\mathbf{v}\wedge\mathbf{B})_{EW}|+c\,|\sin\theta\,\sin\phi|)$
as a function of $c$ and for several values of $|(\mathbf{v}\wedge\mathbf{B})_{EW}|$.
The blue line represents the whole sample of events. The stars located
at $c=10$ give the value of the SD without correction.}
\label{VXBsinSD}
\end{figure}

The observed variations in function of $|(\mathbf{v}\wedge\mathbf{B})_{EW}|+c$ suggest strongly that the radio signal is made of the mixture of several emission effects. Because the observable $\epsilon_0$ seemed sensitive enough to different contributions, its association to an existing electric field was then studied, for instance to an electric field component whose orientation would be linked to the radio shower axis. To address this possibility, we changed the term $c$ by: $\varepsilon_{0}\sim E|(\mathbf{v}\wedge\mathbf{B})_{EW}|+E~c|\sin\theta\,\sin\phi|$. 
The effect of this correction is depicted fig. \ref{VXBsinSD}, which presents
the standard deviation of $(E_{P}-E_{0})/E_{P}$ as a function of $c$ and for the same windows as in $|(\mathbf{v}\wedge\mathbf{B})_{EW}|$. A worse standard deviation is obtained, which leads to exclude the scenario of a longitudinal contribution, relative to the shower axis. 

Some recent developments \cite{mar11, vri11, wer12} could however give room to the simple correction with $c$. For instance the charge excess mechanism could be  in qualitative agreement with the presented effects. Indeed, this emission
depends on the charges produced in the shower (and so from the energy) and its weight in the total emission process should increase when the geomagnetic contribution decreases. The effect of variation of the optical index is also expected to make a contribution rather independent of the geomagnetic angle, although our data do not suggest this contribution in the RLDFs. On another view, our result could also suggest a coherence effect modulated by the Lorentz force. The coherence mechanism should increase when the secondary particles are less dispersed by the Lorentz force (i.e.
at small $|(\mathbf{v}\wedge\mathbf{B})_{EW}|$) during
their propagation. 

\section{Source reconstruction}

Due to its potential link with the quantity $X_{max}$, the reconstruction of an apparent point source in the emission could be one of the significant key-point in the development of radio-detection \cite{laf10}. Because the arrival timing on each antenna is determined at the maximum amplitude of the radio signal, itself dominated by the maximum of the shower \cite{mar112}, this observable should therefore be linked to this  part of the longitudinal development of the shower. The assumption of a spherical wavefront, then gives, in a natural way, a efficient tool to find the position of this instant of emission. Moreover, the self-trigger technique in radio can be subject to delicate limitations in regard to UHECR recognition, due to noises source interferences (human activities, stormy conditions). In this context, the identification of the positions of the sources of emission is crucial. For sources outside of the setup, this identification can be only achieved by using curved wavefront.

\begin{figure}
\centering
\includegraphics[width=8cm,height=6cm]{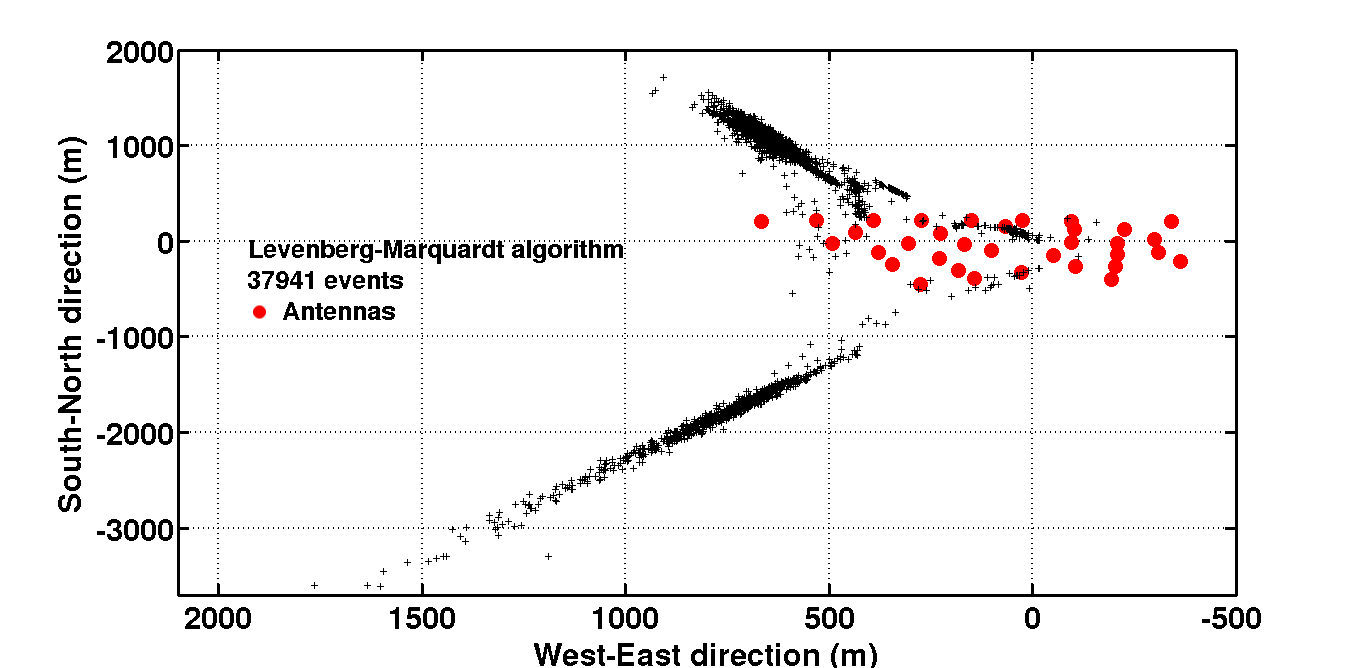}
\caption{Typical result of reconstruction of two entropic emitters at ground, observed with the stand-alone stations of CODALEMA (red circles), through a standard
minimization algorithm. Despite the spreading of the positions reconstructed, both actual transmitters are stationary point sources. Such patterns are also commonly observed in others radio experiments \cite{mohr11}.}
\label{codanoise} 
\end{figure}

For a spherical wavefront, the commonly used technique to extract this curvature radius relies on the calculation of the $\chi^2$ function, using the arrival times and the locations of the antennas, by minimizing the quantity: $\chi^2(\vec{r_{s}},t_{s})=\sum_{i}\left[\left\Vert \overrightarrow{r_{s}}-\overrightarrow{r_{i}}\right\Vert ^{2}-c^2\left(t_{s}-t_{i}\right)^{2}\right]^{2} $ 
with $(\overrightarrow{r_{i}},t_{i})$ the location and reception time of  the $i^{th}$ antenna; $(\overrightarrow{r_{s}},t_{s})$ the searched location and unknown instant of emission of the emitting source; and c the wave velocity considered constant. 
Figure \ref{codanoise} shows a typical reconstruction of sources obtained with CODALEMA, by invoking such a minimization of a spherical wave with a Levenberg-Marquardt (LVM) algorithm. 

\begin{figure}
\centering
\includegraphics[width=8cm,height=6cm]{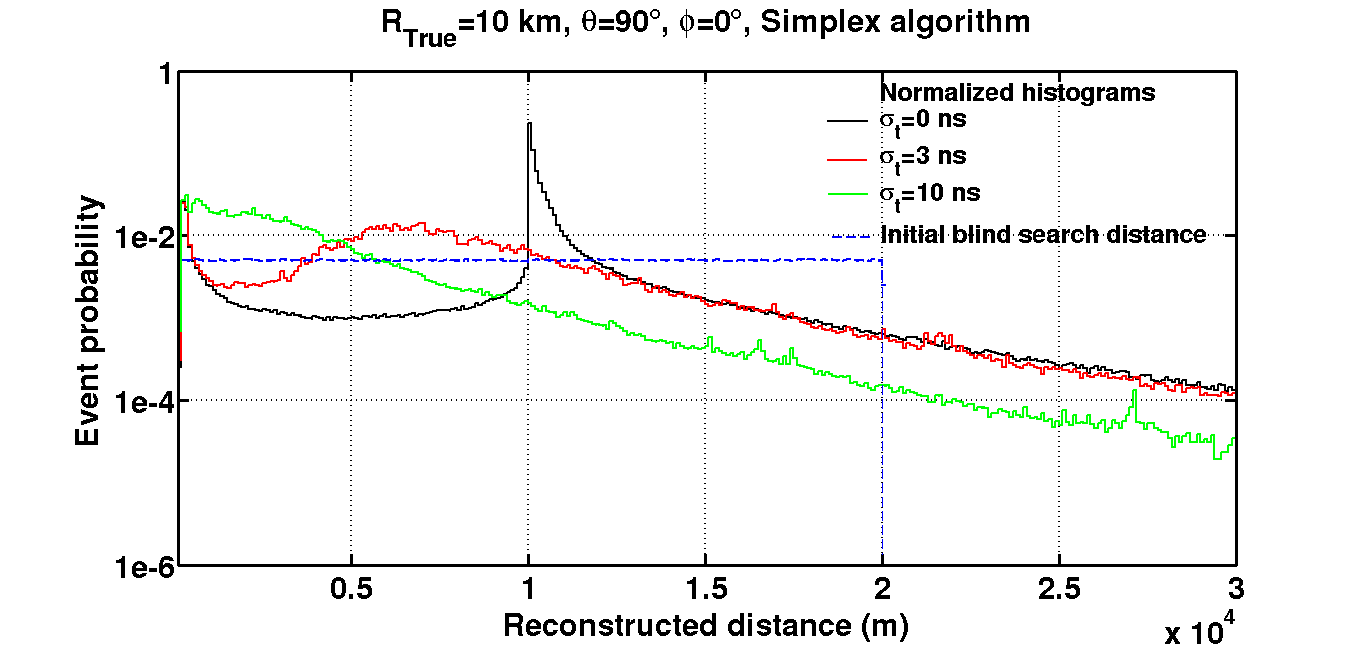}
\caption{Results of the reconstruction, using a testing array of 5 antennas $(400\times 170~m)$, of a source with a radius of curvature of 10 km and the Simplex algorithm. The blue line shows the distribution of the initial values for the blind search.}
\label{Simpl_10km} 
\end{figure}

The source positions are poorly reconstructed, although they are fixed and that the number of events is high. Additional simulations showed that whatever the samples (versus any distances and arrival directions of source, time resolutions, initial conditions, and  detector configurations), different algorithms (LVM, Simplex, Line-search) lead to similar conclusions (see fig. \ref{Simpl_10km}). It suggests this functional seem ill-adapted to the common minimization algorithms. \\

Applied to a cosmic event, which is a single realization of the observables detected (arrival times and peak amplitudes of antennas), such drawbacks could still worsen, even by extracting mean values by using statistical approaches. 
To understand the observations, we have undertaken to study the main features of this $\chi^2$ function \cite{bon03}. \\
 
\begin{figure}
\centering
\includegraphics[width=8cm,height=6cm]{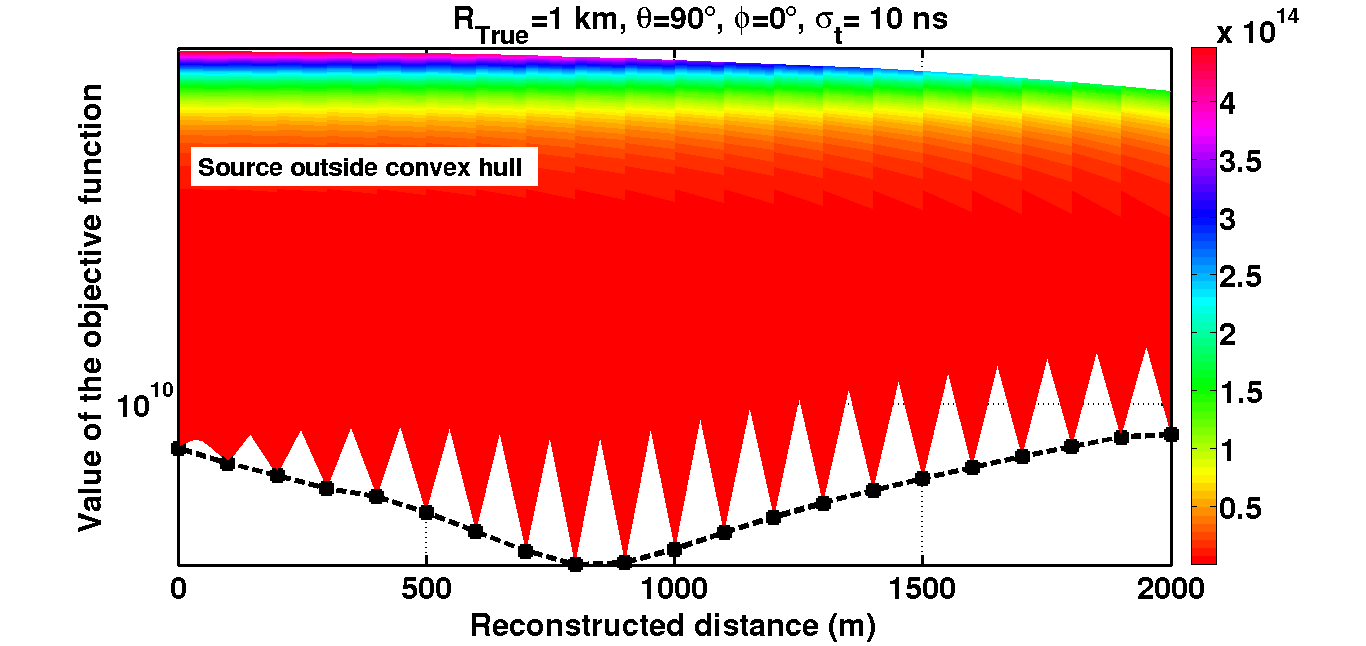}
\caption{Plotting of the objective-function versus $|\vec{r_s}|$ in the case the same testing array as fig. \ref{Simpl_10km} for a source outside the convex hull. This configuration leads to multiple local minima, located on a line joining the antenna barycenter to the true source.}
\label{S_outside}
\end{figure}

Using the criterion of Sylvester \cite{horn85} and the analysis
of the principal minors of the Hessian matrix, we found that the $\chi^2$ function is not convex on small domains, and thus is likely to exhibit several local minima, according to source position relative to the antenna location (fig. \ref{S_outside}). These minima induce problems to converge to the correct solution.\\

The development of the first optimality condition (Jacobian = 0) allowed to study the critical points. It indicates that the iso-barycenter of the antenna array (of the lit antennas for a given event) plays an important role in explaining the observed numerical degeneration. There are strong indications, in agreement with the experimental results and our simulations, that the critical points are distributed on a line connecting this barycenter and the actual source location. In this case the problem is ill-posed.\\ 

\begin{figure}
\centering
\includegraphics[width=8cm,height=4cm]{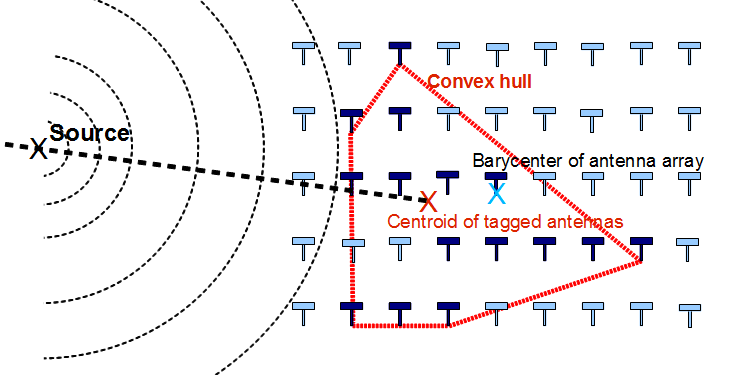}
\caption{Sketch of the topology of a convex hull of a 2D array of antennas for a particular event. The dark markers are the lit antennas for this event. The red dot line shows the corresponding convex envelope. The red cross depicts the centroid of the lit antennas, the black cross represents the barycenter of the antenna array.}
\label{convex}
\end{figure}

This approach has highlighted the role of the convex hull of the array of antennas, which is defined, event by event, by the envelope of the tagged antenna (fig. \ref{convex}). We found that if the source is in the convex hull of the detector, the solution is unique. In contrast, the location of the source outside the convex hull of the detector, causes degeneration of solutions (multiple local minima) regarding to the constrained optimization problem.\\

\begin{figure}
\centering
\includegraphics[width=8cm,height=6cm]{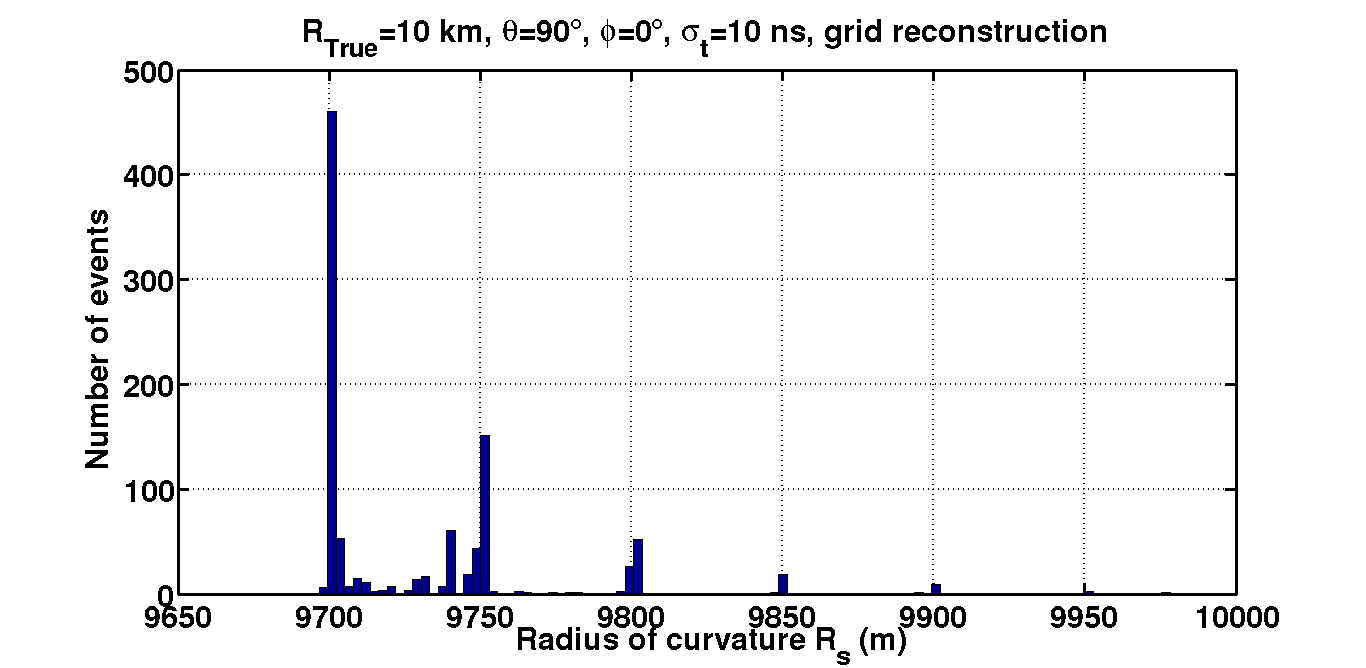}
\caption{Histogram of the reconstruction of a source located at 10 km from the same 
the detector array as in fig. 8, and using the grid method associated to the search
of the global minimum for each event. }
\label{Mindemin}
\end{figure}

We have used this observations to construct
an alternative method of locating the source. To avoid the
trap of the local minima with common algorithms, we chose to compute
directly the values of the objective function on a grid, using a subset
of phase space in the vicinity of the solution a priori, and assuming
that the minimum of the objective function corresponds to the best
estimate of the position of the source of emission. 
The direction-priori is given by the planar
fit, while the quantity $r_{s}$ is left free in the range $0.1-20~km$
(the upper bound being determined by the value of the curvature exploitable,
given the time resolutions currently available). A typical result obtained with our method is presented in fig. \ref{Mindemin}. The first results indicate better accuracies. Next step of this study will consist to exploit all the available data (events other than "internal"), and quantify the bias introduced by this method. 

\section{Conclusion} 

The possibility to extract in radio signal, a simple observable providing an estimate of the energy of the shower has been presented. The maximum amplitude of the radio electric field on the ground ($\epsilon_0$) seems relevant (and surprisingly very usable when compared to particles detectors). We show that this  observable is sensible to different mechanisms causing the radio signal. We demonstrate that the only correction by the geomagnetic effect would lead to an over-evaluation of the energy of the EAS. We deduce that a further contribution is also at the origin of the formation of the radio signal. In a first interpretation, effects like coherence, charge excess, or Cerenkov emission could play a significant role in the observed amplitude modulation. More data should improve obviously the proposed approach, but given the results, the radio signal may be sufficiently sensitive to the longitudinal evolution of the shower, to contain valuable information on the nature of the primary.\\
In second part, we addressed the problem of the apparent point of emission in radio, by reconstructing spherical wavefronts. Indeed, the experimental results indicated that the usual methods of minimization of spherical wavefronts could induce a mis-localization of the emission sources. The specific characteristics of the function $\chi^2$ can lead to the formulation of an ill-posed problem, in the sense of minimization theories under constraints. The conventional algorithms can exhibit a degeneration of the solutions, due to the presence of several local minima. To avoid possible traps in these minima, we tried a simple method. It is based on a grid calculation of the function $\chi^2$ and the search for an absolute minimum in a phase space fixed a priori. Further developments and tests will undoubtedly still needed to use the method routinely. \\

The authors acknowledge the support of the French ``Agence Nationale de la Recherche'', under grant ANR-NT05-2-42808, and the support of the ``Région Pays de la Loire'', under grant 2007/10135.


\end{document}